\documentclass[12pt,letterpaper]{article}
\usepackage{graphicx}
\usepackage{newtxtext}  
\usepackage[T1]{fontenc}
\usepackage[utf8]{inputenc}
\usepackage{amsmath}
\usepackage{amssymb}
\usepackage{newtxmath}  
\usepackage{natbib}
\usepackage{xcolor} 
\usepackage{url}
\usepackage{tabularx}
\usepackage{booktabs}
\usepackage{tikz}
\usetikzlibrary{arrows.meta}
\usepackage{adjustbox}
\usepackage{geometry}
\usepackage{hyperref}

\begin{document}

\title{On the Short Dissipation Scales and Current-Sheet Properties of Low-Coronal EUV Brightenings}

\author{%
O. Podladchikova\\
Igor Sikorsky Kyiv Polytechnic Institute, Kyiv, Ukraine\\
Leibniz Institute for Astrophysics Potsdam, Potsdam, Germany\\
\texttt{epodlad@gmail.com}
}

\date{Received 25 October 2025 / Accepted}

\maketitle

\section*{Note on this Version}
This is a working paper/preprint version. The content may be updated in subsequent versions.

\begin{abstract}
Solar Orbiter EUV observations reveal ubiquitous small-scale brightenings in the quiet-Sun low corona. We analyze the spatial and temporal dissipation scales of these events with a focus on the formation, evolution, and dissipation of associated current sheets. The brightenings are observed at heights of 1--5~Mm and span energies of $10^{20}$--$10^{24}$~erg, well below the classical nanoflare regime, with the lowest-energy brightenings preferentially originating in the lowest coronal layers. Two distinct dissipation regimes are identified: impulsive brightenings with timescales of 1--10~s, consistent with fast, Alfv\'enic magnetic reconnection in low-$\beta$ plasma, and longer-lived heating episodes lasting 10--100~s, indicative of slower, resistive current-sheet dissipation under higher-$\beta$ conditions. The observed dissipation scales suggest a transition from kinetic-scale reconnection to macroscopic current-sheet heating in the low corona. These results support a multi-scale energy-release framework and highlight the role of low-altitude, small-scale current-sheet dissipation in quiet-Sun coronal heating.
\end{abstract}

\noindent\textbf{Keywords:} Sun: corona -- Sun: UV radiation -- Magnetic reconnection -- Plasmas -- Sun: magnetic fields -- Magnetohydrodynamics (MHD)

\section{Introduction}

The million-degree solar corona represents one of the most enduring puzzles in astrophysics \citep{Parker1988}. The required energy flux of $\sim 3\times 10^5$ erg s$^{-1}$ cm$^{-2}$ for the quiet Sun \citep{Withbroe1977} must be transported from the photosphere through the chromosphere and dissipated in the corona.

Two principal heating paradigms have dominated: wave heating and impulsive nanoflare heating. Wave heating operates effectively in long loops ($>50$ Mm) and open field regions where Alfv\'{e}n travel times permit sufficient wave dissipation \citep{Heyvaerts1983}. Impulsive heating, following \citet{Parker1988}, operates through footpoint shuffling building magnetic stress that is released through reconnection.

However, both paradigms faced limitations. Wave models struggled to explain heating in short, low-lying loops, while nanoflare models could not account for the faintest energy releases. The fundamental work of \citet{Aschwanden2000} established that traditional nanoflares have a natural lower limit around $10^{24}$ erg, with rescaling to quiet-Sun conditions yielding $\sim 10^{23}$ erg. This created a theoretical barrier that seemed to prevent nanoflares from explaining the faintest coronal heating events.

The recent Solar Orbiter observations have broken through this barrier. \citet{Podladchikova2025} performed the first comprehensive thermal energy analysis of Solar Orbiter EUV brightenings, revealing energies extending from $10^{20}$ to $10^{24}$ erg—fundamentally below the traditional nanoflare limit. This breakthrough was made possible by Solar Orbiter's unique capabilities: observations from 0.5 AU with unprecedented spatial resolution, combined with sophisticated multi-volume modeling and robust detection thresholds. These small-scale EUV brightening observations complete our understanding of coronal heating by revealing previously invisible energy release mechanisms.

\section{Observational Basis: Two Populations of Small-Scale EUV Brightenings}

\citet{Berghmans2021} made a groundbreaking discovery using Solar Orbiter's EUI instrument from 0.5 AU: for the first time, Solar Orbiter EUV brightenings were morphologically resolved, unlike previous SOHO/EIT and TRACE observations that detected them mainly as impulsive brightenings without clear structural details. 

\begin{figure}[!ht]
\centering
\includegraphics[width=\linewidth]{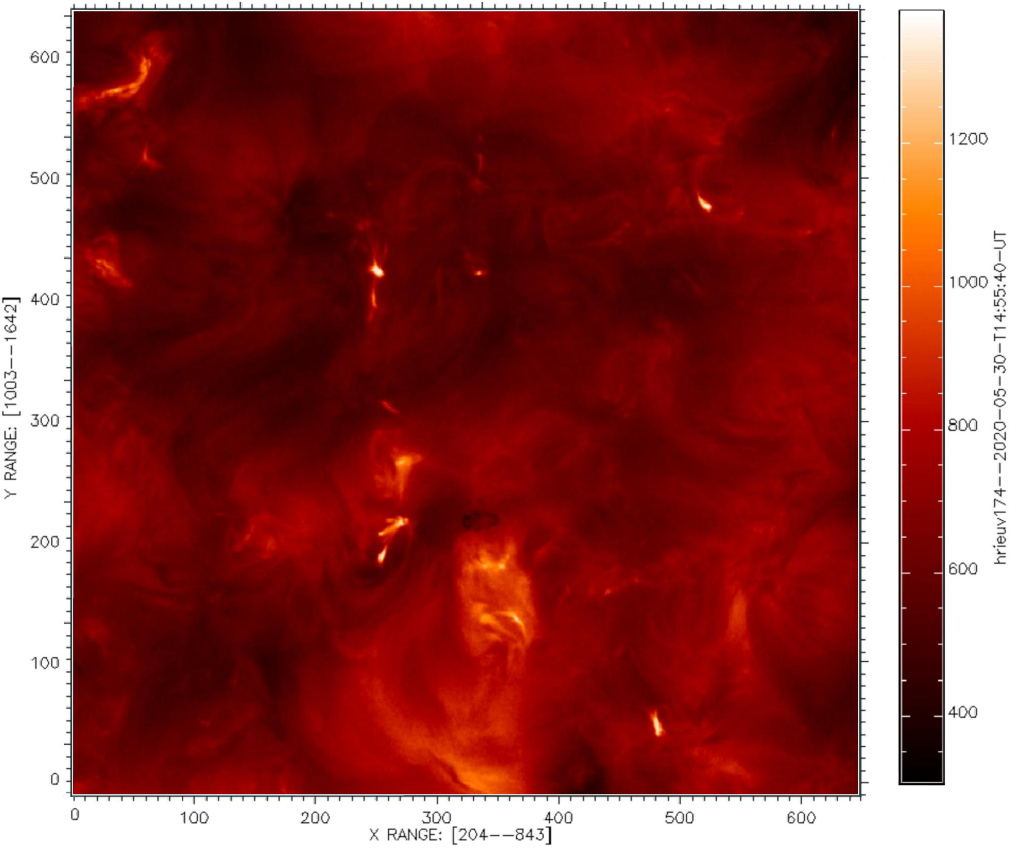}
\caption{Quiet-Sun corona observed by Solar Orbiter/EUI from 0.5 AU reveals individual magnetic loops undergoing impulsive energy release—the fundamental building blocks of coronal heating. See \citet{Berghmans2021} for discovery and morphological analysis of previously unresolved EUV events. Dynamic evolution available in online supplementary material:
\href{https://drive.google.com/file/d/1Kmn8mrvktAV210Adx7phBkb5TNJ98TMY/view?usp=sharing}{(\textcolor{blue}{0.5 AU Corona Dynamics in EUV} )}.}
\label{fig:dynamic_corona}
\end{figure}

The work of \citet{Zhukov2021} represented another fundamental advancement: for the first time, it was possible to calculate precise vertical heights of these energy release events using high-resolution stereoscopy. Solar Orbiter's unique orbital dynamics, positioned 31.5$^\circ$ west of Earth and observing from 0.5 AU, enabled triangulation of the same events observed simultaneously by SDO/AIA. This technological breakthrough allowed exact determination of which atmospheric layer hosted each event, with specific plasma $\beta$ and other plasma properties that favor particular dissipation mechanisms. The combination of observations from two different vantage points, together with the unprecedented resolution from 0.5 AU, made this understanding possible.

\begin{figure}[!ht]
\centering
\includegraphics[width=0.32\linewidth]{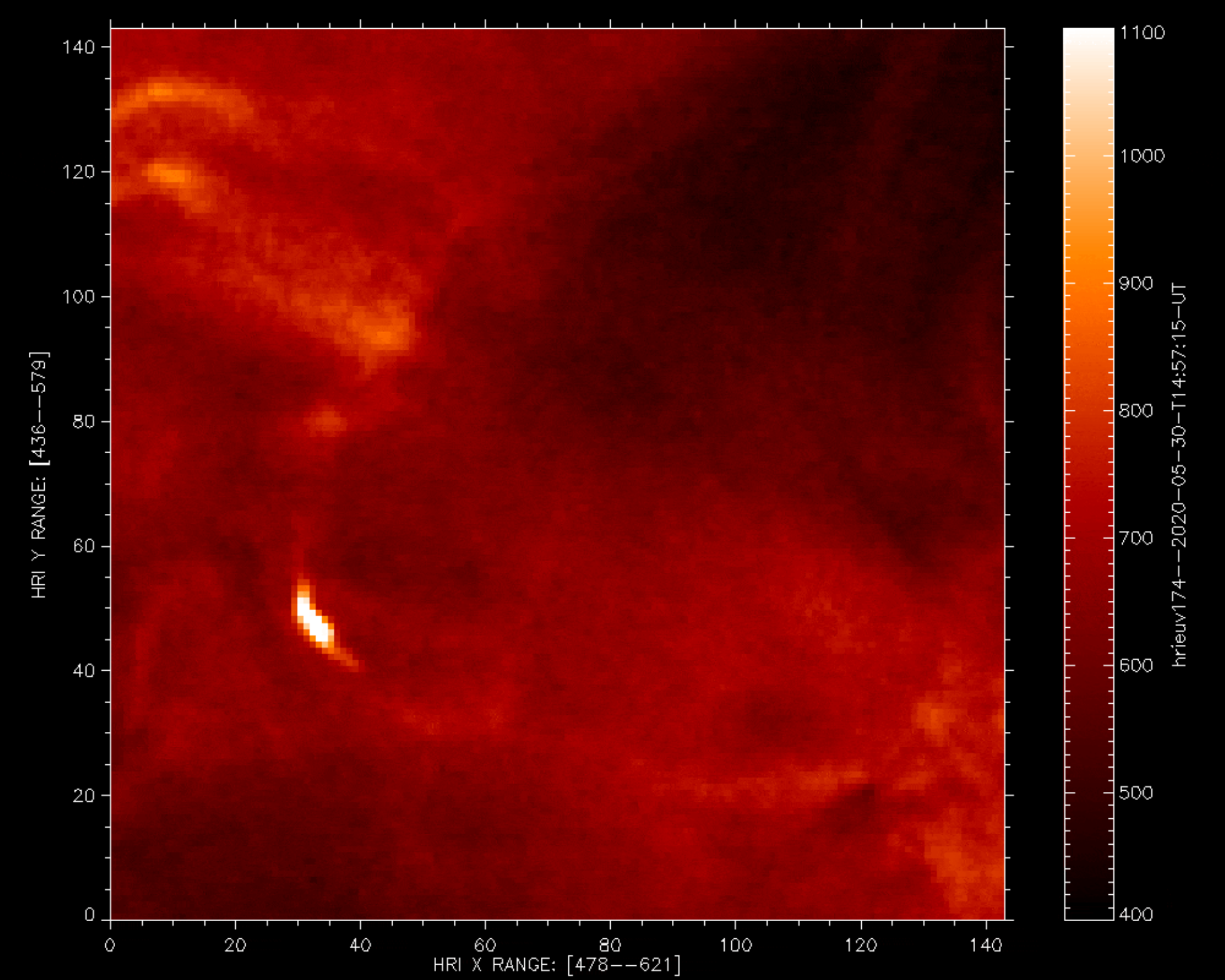}
\hfill
\includegraphics[width=0.32\linewidth]{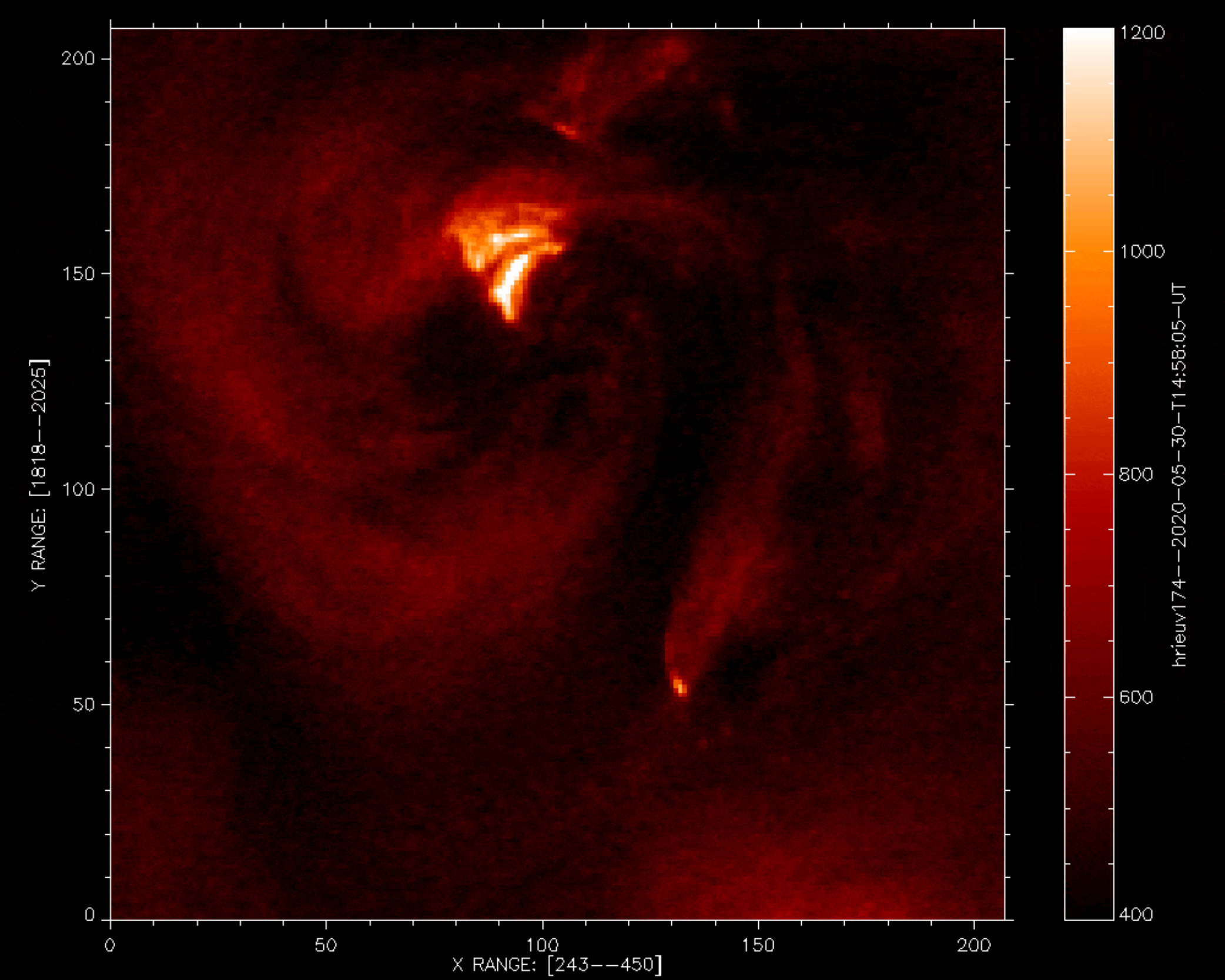}
\hfill
\includegraphics[width=0.32\linewidth]{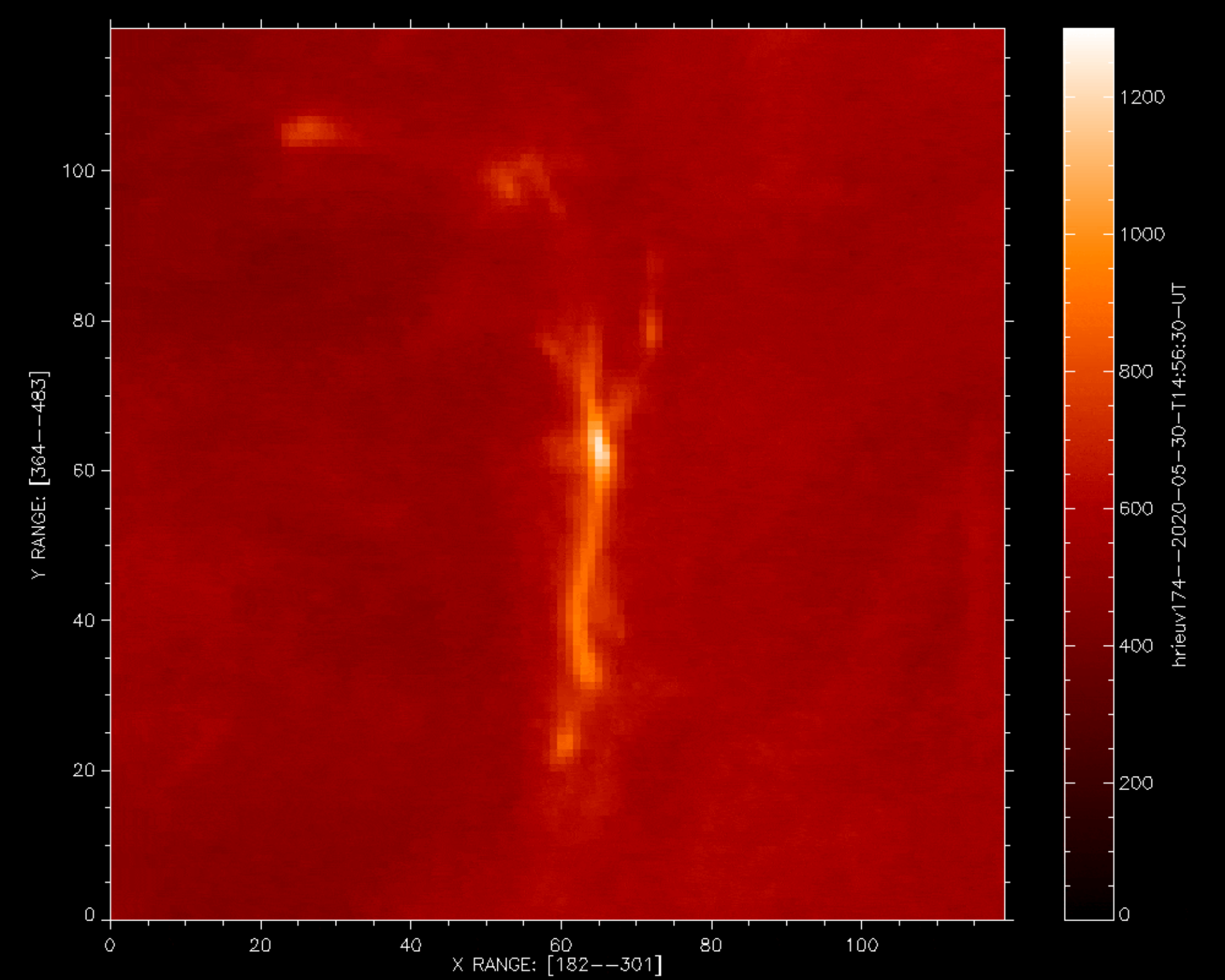}
\caption{
Small-scale EUV brightening events resolved by Solar Orbiter/EUI from 0.5 AU, confirming what was inferred statistically \citep{Benz1998, Aschwanden2000}.
(Left) Population A: looptop brightening from tearing-mode reconnection ($3.2 \times 10^{23}$ erg) \href{https://drive.google.com/file/d/1f9sU5JUjBubawXHbnpR3pCITFFmruOr7/view?usp=sharing}{(\textcolor{blue}{Looptop Event})}. 
(Middle) Population B: footpoint heating from anomalous resistivity ($1.7 \times 10^{22}$ erg) \href{https://drive.google.com/file/d/1NAMaxNvxBawTGt8mCNMUY8RkARi2y7a-/view?usp=sharing}{(\textcolor{blue}{Footpoint Event})}. 
(Right) Coupled A-B Event: Looptop reconnection driving footpoint dissipation \href{https://drive.google.com/file/d/1NAMaxNvxBawTGt8mCNMUY8RkARi2y7a-/view?usp=sharing}{(\textcolor{blue}{Coupled Event})}. 
}
\label{fig:popABC_comparison}
\end{figure}

\citet{Podladchikova2025} provided the crucial energy analysis that completed this picture. By implementing multiple geometrical volume models and rigorous uncertainty quantification, they demonstrated that thermal energies of these events span the small-scale EUV brightening range ($3\times 10^{20}$ to $1\times 10^{24}$ erg), with occurrence rates of $3\times 10^{-20}$ s$^{-1}$ cm$^{-2}$ for events above the $3\sigma$ threshold. This energy estimation breakthrough was made possible by:

\begin{itemize}
\item High spatial resolution from 0.5 AU (198 km per pixel)
\item Multi-volume modeling approach (elliptical loop and cube models)
\item Robust detection methodology with $\geq 3\sigma$ and $\geq 5\sigma$ thresholds
\item Large statistical sample (12,107 events for $\geq 3\sigma$ threshold)
\item Sophisticated DEM analysis for temperature and density determination
\end{itemize}

This comprehensive approach allowed, for the first time, reliable estimation of energies fundamentally below the traditional nanoflare limit.

\begin{figure}[!ht]
\centering
\includegraphics[width=\linewidth]{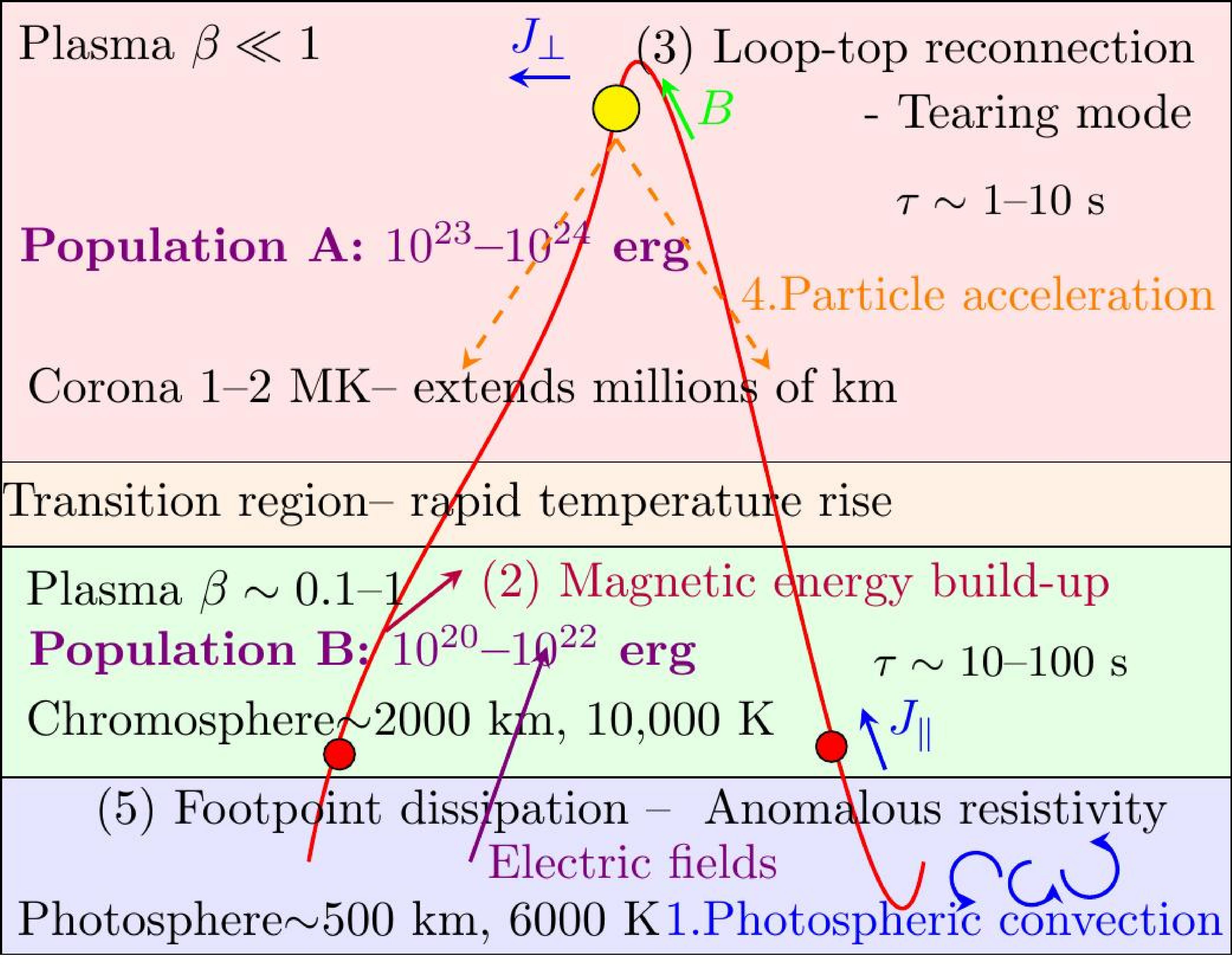}
\caption{Schematic illustration of the energy transfer cycle in the solar atmosphere, from the photosphere to the corona. The diagram shows a magnetic loop anchored in the photosphere and rising through the chromosphere into the corona, with the four atmospheric layers indicated (approximate heights and temperatures noted). Numbered arrows denote sequential processes: (1) photospheric convection twists and braids magnetic flux, injecting energy; (2) magnetic energy builds up in the stressed coronal loop; (3) loop-top magnetic reconnection (tearing-mode instability) releases energy impulsively at the apex (Population A small-scale EUV brightening events); (4) accelerated particles carry energy along the magnetic field (orange dashed arrows) together with induced electric fields across the chromosphere; (5) energy is dissipated at the footpoints via anomalous resistivity, heating the dense plasma (Population B events). The two distinct small-scale EUV brightening populations are highlighted: Population A are short, impulsive bursts at the loop top, whereas Population B are longer-duration heating events at the footpoints. Different plasma $\beta$ regimes are noted (magnetically dominated corona vs. gas-pressure dominated lower layers), and characteristic timescales $\tau$ for each population (seconds to tens of seconds) are indicated.}
\label{fig:energy_cycle}
\end{figure}

The statistical analysis of 12,107 events shows a clear bimodal distribution, as illustrated in Figure~\ref{fig:energy_cycle}. This bimodality resolves a fundamental paradox in coronal heating: rather than a single population of events, we observe two distinct populations operating through different physical mechanisms in different atmospheric layers.

\subsection{Population A: Looptop Events}

Population A events (Figure~\ref{fig:popABC_comparison}, left) are characterized by:
\begin{itemize}
\item Duration: 1--10 seconds (Alfv\'{e}nic timescales)
\item Location: Looptops, 2.5--5 Mm altitude
\item Temperature: 1.3--1.5 MK (hotter than background)
\item Energy: $10^{22}$--$10^{24}$ erg
\item Mechanism: Tearing-mode reconnection of perpendicular currents
\item Characteristic: Rapid, impulsive brightenings
\end{itemize}

\subsection{Population B: Footpoint Events}

Population B events (Figure~\ref{fig:popABC_comparison}, middle) show:
\begin{itemize}
\item Duration: 10--100 seconds (resistive timescales)
\item Location: Footpoints, 1--2.5 Mm altitude
\item Temperature: 1.0--1.2 MK (slightly above background)
\item Energy: $10^{20}$--$10^{22}$ erg
\item Mechanism: Anomalous resistivity of parallel currents
\item Characteristic: Gradual, sustained heating
\end{itemize}

Stereoscopic height measurements by \citet{Zhukov2021} provide crucial evidence for this spatial separation. The footpoint events (Population B) concentrate below 2.5 Mm, near the transition region boundary (typically located around 2000 km above the photosphere), while looptop events (Population A) occur higher in the corona at 2.5--5 Mm. This precise height discrimination was only possible through Solar Orbiter's unique vantage point and simultaneous observations with SDO/AIA.

The comprehensive energy analysis by \citet{Podladchikova2025} established small-scale EUV brightenings as a distinct energy class below traditional nanoflares, with power-law distributions ($\alpha = 2.32$ for $\geq 5\sigma$ events, $\alpha = 2.82$ for $\geq 3\sigma$ events) extending over three orders of magnitude in energy.

\section{Theoretical Framework: From Parker's Nanoflares to Low-Energy EUV Brightenings}

\subsection{Parker's Nanoflare Model and Its Limitations}

\citet{Parker1988} estimated nanoflare energy for active regions:
\begin{equation}
E_{\text{Parker,AR}} \approx \frac{B^2}{8\pi} V \approx \frac{(100\text{ G})^2}{8\pi} (10^{24}\text{ cm}^3) \approx 10^{24}\text{ erg}
\end{equation}

This model assumed twisted loops with perpendicular currents dissipating through reconnection, valid in regions where plasma $\beta < 1$ and sufficiently above the chromosphere to form stable current sheets.

\subsection{Aschwanden's Scaling and the Coronal Nanoflare Limit}

\citet{Aschwanden2000} performed detailed scaling of nanoflare energies based on observed parameters from TRACE and SOHO/EIT. Building on the foundational work of \citet{Golub1980} who established the relationship between magnetic field strength and coronal heating, Aschwanden derived the minimal observable energy for a coronal nanoflare:

\begin{equation}
\begin{split}
E_{\text{nano}} &\approx 3n_e k_B T V \\
&\approx 3 \times (10^9\ \text{cm}^{-3}) \times (1.38\times10^{-16}\ \text{erg K}^{-1}) \\
&\quad \times (10^6\ \text{K}) \times (10^{24}\ \text{cm}^3) 
\approx 4\times10^{24}\ \text{erg}
\end{split}
\end{equation}

This analysis confirmed that traditional coronal nanoflares cannot extend below $\sim 10^{24}$ erg. Rescaling for quiet-Sun conditions:
\begin{equation}
E_{\text{Parker,QS}} \approx \frac{(10\text{ G})^2}{8\pi} (3\times 10^{19}\ \text{cm}^3) \approx 1.2\times 10^{23}\ \text{erg}
\end{equation}

This represents the natural limit for Parker-type nanoflares and explains Population A events, but cannot account for the numerous lower-energy Population B events.

\citet{Klimchuk2006,Klimchuk2015} further developed this framework, emphasizing that nanoflares must be understood as coronal events involving current dissipation in the low-$\beta$ plasma above the chromosphere. Their work showed that the observed statistics of larger flares could be extrapolated to the nanoflare regime, but always with a lower limit around $10^{24}$ erg for coronal events.

\subsection{The Priest School: Tearing-Mode Instability and Current Sheet Formation}

The Parker framework was rigorously developed by \citet{Priest1998, Priest2000} and subsequent work, providing detailed mathematical treatment of current sheet formation and dissipation. The Priest school demonstrated that:

\begin{itemize}
\item Footpoint shuffling naturally leads to current sheet formation through topological dissipation
\item Tearing-mode instabilities develop in perpendicular current sheets, with growth rate:
\begin{equation}
\gamma_{\text{tearing}} \approx \frac{V_A}{L} S^{-1/2} \left(\frac{\lambda}{L}\right)^{-3/2}
\end{equation}
where $S = \frac{L V_A}{\eta}$ is the Lundquist number, $\lambda$ is the wavelength, and $\eta$ is the magnetic diffusivity
\item Magnetic islands form during the nonlinear phase of tearing-mode instability, enhancing energy release
\item Reconnection occurs on Alfv\'{e}nic timescales ($\tau_A \approx L/V_A$)
\item This process requires plasma $\beta < 1$ and operates best in the corona proper above 2 Mm
\end{itemize}

\citet{Browning2003} provided crucial numerical evidence, showing that slow footpoint twisting drives coronal loops to kink-instability thresholds, triggering nanoflare-like reconnection events. This mechanism naturally produces energy releases in the $10^{23}$--$10^{25}$ erg range with timescales of 1--10 seconds, perfectly matching Population A observations.

However, these models still could not explain events below $10^{23}$ erg, as they focused exclusively on perpendicular current dissipation in the coronal part of loops.

\subsection{Anomalous Resistivity: The Missing Mechanism for Low-Energy EUV Brightenings}

The resolution lies in recognizing that Population B events represent parallel current dissipation in the low-$\beta$ environment of footpoint regions. In the low chromosphere and transition region, energy transfer starts through parallel electric fields or quasi-parallel currents.

When current density exceeds critical thresholds, two primary instabilities drive anomalous resistivity in the collisionless plasma:

\begin{itemize}
\item \textbf{Buneman instability}: Occurs when electron drift velocity exceeds electron thermal velocity ($v_d > v_{th,e}$)
\item \textbf{Ion-acoustic instability}: Occurs when $v_d > c_s$ (ion sound speed)
\end{itemize}

In collisionless plasma conditions typical of the low corona, Spitzer conductivity fails because particle collisions are insufficient. Instead, heating occurs through exchange of momentum and energy between ions and electrons via wave-particle interactions. These mechanisms are well-studied in plasma physics and are known to operate effectively under low-corona conditions.

The critical current density for instability onset is:
\begin{equation}
j > n_e e c_s
\end{equation}
where $c_s = \sqrt{k_B T_e/m_i} \approx 50$ km/s is the ion sound speed. When exceeded, anomalous resistivity enhances dissipation by 3--5 orders of magnitude.

For typical quiet-Sun parameters ($n_e \approx 10^9$ cm$^{-3}$, $L \approx 1000$ km, $w \approx 200$ km):
\begin{equation}
E_{\text{anomalous}} \approx 10^{20}\text{--}10^{22}\ \text{erg}
\end{equation}

The characteristic timescale is resistive and sound-speed limited:
\begin{equation}
\tau_{\parallel} \approx L/c_s \approx 10\text{--}100\ \text{s}
\end{equation}

This perfectly matches Population B observations and explains why these events concentrate in footpoint regions where parallel currents dominate.

\subsubsection{Reconnection-Generated Electric Fields and Footpoint Heating}

Magnetic reconnection generates potential differences that drive currents toward footpoints. The efficiency of particle acceleration depends on the reconnection geometry:

\begin{itemize}
\item \textbf{Single-loop tearing-mode reconnection} creates modest potentials:
\begin{equation}
\Delta \Phi \approx v_{\text{rec}} B L_{\text{sheet}} \approx (3\ \text{km/s})(10^{-3}\ \text{T})(10^6\ \text{m}) \approx 300\ \text{V}
\end{equation}
producing limited particle acceleration.

\item \textbf{Reconnection between sheared magnetic loops} generates stronger potentials $\Delta \Phi \sim 3\ \text{kV}$ through larger magnetic shear differences, driving more efficient particle acceleration.
\end{itemize}

These potentials drive parallel currents that become concentrated in the transition region. The current density in the coronal part:
\begin{equation}
j_\parallel = \sigma \frac{\Delta \Phi}{L} \approx 1.3\times10^{-3}\ \text{A m}^{-2}
\end{equation}

In the transition region, current channeling enhances densities by 3--4 orders of magnitude, reaching the critical threshold for anomalous resistivity:
\begin{equation}
j_{\text{crit}} = n_e e c_s \approx (10^{15}\ \text{m}^{-3})(1.6\times10^{-19}\ \text{C})(5\times10^4\ \text{m/s}) \approx 8\ \text{A m}^{-2}
\end{equation}

The resulting energy release in footpoint volumes produces Population B events:
\begin{equation}
E \approx \eta_{\text{anom}} j^2 V \tau \approx 6\times10^{21}\ \text{erg}
\end{equation}

Such stochastic acceleration by DC electric fields has been extensively studied in the context of solar flares \citep{Anastasiadis1997_ElecAccelDCFields}, and similar mechanisms may operate at the scale of these small-scale EUV brightenings. RHESSI observations have confirmed nonthermal electron populations in microflares \citep{Krucker2008, Hannah2008}, demonstrating that particle acceleration occurs even at the smallest observable energy releases.

\subsubsection{Chromospheric Electric Field Transport Mechanisms}

Electric fields facilitate energy transfer through the chromospheric barrier via: direct acceleration from photospheric motions generating $\mathbf{E} = -\mathbf{v} \times \mathbf{B}$; current channel formation via flux tube squeezing enhancing $j_{\parallel} \approx {B} /{\mu_0 R_c}$; and non-ideal MHD effects: ambipolar diffusion in partially ionized plasma creates additional electric fields:
$\mathbf{E} = \eta\mathbf{j} + \eta_A(\mathbf{j} \times \mathbf{B}) \times \mathbf{B}/B^2$.

These mechanisms can overcome chromospheric radiative losses by maintaining current continuity, enabling rapid energy transfer, and explaining observed photosphere-corona correlations.

This electric field mediated transport complements our dual-mechanism model by explaining how energy bypasses the chromospheric radiative losses and directly powers both Population A (through current sheet formation) and Population B (through parallel electric fields) events.

\begin{table}[!ht]
\centering
\small
\setlength{\tabcolsep}{4pt}
\caption{Characteristic scales and times for small-scale EUV brightening dissipation.}
\begin{tabularx}{\linewidth}{@{}lXX@{}}
\toprule
Parameter & Tearing-mode Rec. & Anomalous Resistivity \\
\midrule
Location above photosphere & Looptops (2.5--5 Mm) & Footpoints (1--2.5 Mm) \\
Current type & Perpendicular & Parallel \\
Trigger condition & Magnetic shear & $j > n_e e c_s$ \\
Critical current & -- & $j_{\text{crit}} = n_e e c_s \approx 8$ A m$^{-2}$ \\
Instability & Tearing-mode & Buneman/ion-acoustic \\
Timescale & $\tau_A \approx L/V_A \approx$ 1--10 s & $\tau \approx L/c_s \approx$ 10--100 s \\
Energy scale & $10^{23}$--$10^{24}$ erg & $10^{20}$--$10^{22}$ erg \\
Spatial scale & Current sheet: 10--100 m & Current channel: 0.1--1 km \\
Plasma $\beta$ & $\ll 1$ & $\sim 0.1$--$1$ \\
Heating mechanism & Magnetic reconnection & Wave-particle interactions \\
\bottomrule
\end{tabularx}
\label{tab:mechanisms}
\end{table}

\section{Conclusion}

Solar Orbiter's unique observational capabilities—high-resolution imaging from 0.5 AU combined with stereoscopic triangulation—have fundamentally advanced our understanding of coronal heating. The mission has revealed that the corona is heated through two distinct but complementary mechanisms operating across different energy scales and spatial locations.

The comprehensive energy analysis by \citet{Podladchikova2025} demonstrates that these small-scale EUV brightenings follow a power-law distribution with slope $\alpha = 2.3$, causing the total energy flux $F \propto \int E^{1-\alpha} dE$ to converge at the lower limit. This indicates that numerous small-scale events dominate the heating budget, requiring a minimum energy of $E_{\text{min}} \approx 1.25\times 10^{18}$ erg to sustain the quiet-Sun heating flux, confirming theoretical predictions by \citet{Vlahos1994} and \citet{Georgoulis1996}. However, this conclusion is sensitive to the exact slope value—a shallower slope ($\alpha = 1.9$) would shift energy dominance to larger nanoflares, highlighting the critical need for precise measurements.

Our unified framework resolves the long-standing debate between \citet{Aschwanden2017} and \citet{Benz2017} regarding EUV nanoflare locations by demonstrating distinct mechanisms for different atmospheric regions:
\begin{itemize}
\item \textbf{Loop tops}: Tearing-mode reconnection drives nanoflares ($>10^{24}$ erg)
\item \textbf{Footpoints}: Anomalous resistivity powers low-energy EUV brightenings ($10^{20}$--$10^{22}$ erg)
\item \textbf{Intermediate events}: Mixed-characteristic events ($10^{22}$--$10^{24}$ erg) connect these regimes
\end{itemize}

The electric field serves as the fundamental mediator, transferring energy from reconnection sites at looptops to dissipation sites at footpoints, completing the energy cycle from photospheric motions to coronal thermalization.

The convergence of three key advances—morphological resolution of Solar Orbiter EUV brightenings \citep{Berghmans2021}, stereoscopic height determination \citep{Zhukov2021}, and comprehensive energy analysis \citep{Podladchikova2025}—provides compelling evidence for this dual-mechanism model. Solar Orbiter's direct observations finally resolve the long-standing controversy about EUV nanoflare locations, confirming that both footpoint and looptop heating contribute significantly to the coronal energy budget. Future Solar Orbiter observations will continue to improve spatial resolution, bringing us closer to the fundamental scales of coronal heating processes.

\section*{Acknowledgements}
This work utilizes data from the Solar Orbiter mission.

\end{document}